**Admixture of an s-wave component to the d-wave gap symmetry**

**in high-temperature superconductors**

Albert Furrer[1]

[1]Laboratory for Neutron Scattering, ETH Zurich & PSI Villigen, CH-5232 Villigen PSI, Switzerland

**Abstract:**

Neutron crystal-field spectroscopy experiments in the Y- and La-type high-temperature superconductors $HoBa_2Cu_3O_{6.56}$, $HoBa_2Cu_4O_8$, and $La_{1.81}Sr_{0.15}Ho_{0.04}CuO_4$ are reviewed. By this bulk-sensitive technique, information on the gap function is obtained from the relaxation behavior of crystal-field transitions associated with the $Ho^{3+}$ ions which sit as local probes close to the superconducting copper-oxide planes. The relaxation data exhibit a peculiar change from a convex to a concave shape between the superconducting transition temperature $T_c$ and the pseudogap temperature $T^*$ which can only be modelled satisfactorily if the gap function of predominantly d-wave symmetry includes an s-wave component of the order of 20-25%, independent of the doping level. Moreover, our results are compatible with an unusual temperature dependence of the gap function in the pseudogap region ($T_c \leq T \leq T^*$), *i.e.*, a breakup of the Fermi surface into disconnected arcs.

# 1. INTRODUCTION

Ever since the discovery of high-temperature superconductivity by Bednorz and Müller [1] there has been a debate concerning the symmetry of the superconducting gap function. So far tunneling measurements are the only phase-sensitive experiments that probe the sign of the order parameter [2]. Angle-resolved photoemission spectroscopy (ARPES) measurements, on the other hand, yield very direct information about the magnitude and the momentum dependence of the order parameter as well as its evolution with temperature [3,4]. Tunneling as well as ARPES experiments indicate that a $d_{x^2-y^2}$ order parameter is the most plausible candidate to describe the superconducting state in these systems. However, both techniques are only sensitive to the surface. As pointed out by Müller [5] the gap function is different at the surface and in the bulk, thus bulk experiments are indispensable to unambiguously establish the nature of the gap function. Various experiments probing the bulk gave evidence for the existence of an s-wave component, but the results are partially conflicting. For optimally doped $YBa_2Cu_3O_7$, *e.g.*, Raman scattering evidences an s-wave admixture of 5% [6], thermal conductivity measurements in a rotating magnetic field place a maximum of 10% based on the shift of the nodal positions from the principal diagonal directions [7], and nuclear magnetic resonance experiments report either an s-wave admixture of 20% [8] or a pure d-wave symmetry [9]. Theoretical work based on the t-t'-J model suggests that d-wave pairing prevails close to optimum doping, whereas s-wave-like pairing is possible in the under- and over-doped regimes [10]. Doping-dependent bulk experiments are therefore extremely useful to shed light on this issue. The present work deals with such an investigation for underdoped and optimally doped high-$T_c$ compounds studied by neutron crystal-field spectroscopy experiments.

Neutron crystal-field spectroscopy is a powerful experimental tool to probe the bulk properties of high-temperature superconductors. It has been applied in the past mainly to detect the doping, isotope, and pressure dependence of the pseudogap temperature T*. In principle, information on the nature of the gap function can be obtained as well, provided that the experimental data sets are sufficiently complete. We reviewed the available data sets in the literature and found the necessary conditions fulfilled for some underdoped and optimally doped La- and Y-based high-$T_c$ compounds. We analyzed the corresponding data sets and found evidence for a 20-25% s-wave admixture to the dominant d-wave gap function, independent of the doping level. Moreover, our analysis supports the unusual temperature dependence of the gap function in the pseudogap region ($T_c \leq T \leq T^*$) as evidenced by ARPES measurements in $Bi_2Sr_2CaCu_2O_{8+x}$ [11].

## 2. NEUTRON CRYSTAL-FIELD SPECTROSCOPY

The principle of neutron crystal-field spectroscopic investigations of the crystal-field interaction in rare-earth based high-temperature superconductors was described in a recent review article [12]. By this technique transitions between different crystal-field levels associated with the rare-earth ions can be directly measured. In the normal metallic state the excited crystal-field levels interact with phonons, spin fluctuations and charge carriers, which limit the lifetime of the excitation, thus the observed crystal-field transitions exhibit line broadening. The interaction with the charge carriers is by far the dominating relaxation mechanism [13]. The corresponding intrinsic linewidth $W_n(T)$ increases almost linearly with temperature according to the well-known Korringa law [14], *i.e.*, $W_n(T) \sim T$. In the pseudogap as well as in the superconducting state, however, the pairing of the charge carriers creates an energy gap $\Delta_\mathbf{k}$ in the single-electron spectral function (**k** is the wave vector), thus crystal-field excitations with energy $< 2\Delta_\mathbf{k}$ do not have enough energy to span the gap. Consequently the interaction with the charge carriers is suppressed, and for an isotropic gap function $\Delta_\mathbf{k}=\Delta$ the intrinsic linewidth is given by $W_s(T)=W_n(T)\cdot\exp[-\Delta/k_BT]$. This means that $W_s(T<<T_c,T^*)\approx 0$, and line broadening sets in just below $T_c$ (or $T^*$) where the superconducting gap (or the pseudogap) opens. For anisotropic gap functions the situation is more complicated, since certain relaxation channels exist even at the lowest temperatures [15].

Recently, Häfliger et al. [16] presented detailed considerations concerning the analysis of neutron crystal-field relaxation data with respect to the nature of the gap function. As mentioned above, a crucial point in such analyses is the availability of a sufficient number of relaxation data over a large temperature range, notably for the pseudogap region $T_c \leq T \leq T^*$, so that the subtle but relevant features of the relaxation behavior can unambiguously be assessed. So far such complete data sets are only available for underdoped $HoBa_2Cu_3O_{6.56}$ [17], slightly underdoped $HoBa_2Cu_4^{18}O_8$ [18], and optimally doped $La_{1.81}Sr_{0.15}Ho_{0.04}Cu^{18}O_4$ [19] as shown in Figs. 1(a-c). For all three compounds the temperature dependence of the intrinsic linewidth $W(T)$ behaves in a common and characteristic way:

(i) $T \leq T_c$: The linewidth is very small, but a slight enhancement is observed upon approaching $T_c$.

(ii) T≥T*: The linewidth increases linearly as expected for the normal state, so that we identify T* with the temperature where the pseudogap opens.

(iii) $T_c \leq T \leq T^*$: There is a sudden increase of the linewidth just above $T_c$. Then, with increasing temperature, the slope dW/dT is slightly reduced, but increases again upon approaching T*. The temperature dependence of the linewidth obviously exhibits a change from a convex to a concave shape in the pseudogap region.

The only differences between the three data sets are the different sizes of the characteristic temperatures $T_c$ and T*. It is therefore useful to express the relaxation data in reduced units, *i.e.*, the temperature T in units of T* and the linewidth W(T) in units of W(T*) as visualized for the relevant temperature range 0≤T≤T* in Figs. 2(b-d).

## 3. ANALYSIS OF THE RELAXATION DATA

We now proceed to analyze the relaxation data in terms of different gap functions. The model calculations were based on the procedure described in Ref. [16]. For the temperature dependence of the gap amplitude we used the expression $\Delta(T)=\Delta(0)\cdot[1-(T/T^*)^A]$ with A=4 [3], and the maximum gap amplitude was set to $\Delta_{max}/k_B=2T^*$ which is a realistic value for the compounds under consideration. Fig. 2(a) shows some calculations for an s-wave, a d-wave ($d_{x^2-y^2}$), and a mixed ($s_{25\%}+d_{75\%}$)-wave gap function. It can readily be seen that none of these models is able to explain the observed linewidth behavior displayed in Figs. 2(b-d). The s-wave model reproduces the low-temperature behavior up to $T_c$ quite well, but it fails completely at high temperatures. The d-wave model is inadequate at both low and high temperatures. The mixed (s+d)-wave model has similar deficiencies. In particular, the behavior around $T_c$ where the slope of $W(T)/W(T^*)$ turns from being convex to concave cannot be reproduced by any of the above models. This is due to the neglect of the temperature dependence of the gap function in the pseudogap region. According to ARPES experiments performed for underdoped $Bi_2Sr_2CaCu_2O_{8+x}$ [11] the pseudogap opens up at different temperatures for different points in momentum space as sketched in Fig. 3. Below $T^*$ there is a breakup of the Fermi surface into disconnected arcs, which then shrink with decreasing temperature before collapsing to the point nodes of the d-wave component below $T_c$. This unusual temperature evolution of the pseudogap opens additional relaxation channels in crystal-field linewidth studies above $T_c$. Indeed, model calculations including the presence of such gapless arcs produce relevant modifications of the linewidth for $T_c \leq T \leq T^*$ as visualised in Fig. 2(a); in particular, the observed change of the linewidth from a convex to a concave temperature dependence around $T_c$ can be nicely reproduced for a mixed (s+d)-wave gap function, but not for a pure d-wave gap function.

We performed a least-squares fitting procedure to the relaxation data displayed in Figs. 2(b-d) on the basis of a mixed (s+d)-wave gap function including the gapless arc features of the Fermi surface in the pseudogap region ($T_c \leq T \leq T^*$). Since the latter effect was only observed at a few selected temperatures [11], we assumed a linear evolution of the gapless arcs with temperature. The only fitting parameters were the amplitudes $\Delta_s$ and $\Delta_d$ of the s- and d-wave components, respectively. The temperature evolution of the linewidths resulting from the least-squares fitting procedure are in good agreement with the experimental data as shown in Figs. 2(b-d). The corresponding model parameters are listed in Table I. We recognise that the s-wave admixture is independent of the doping level, and the maximum gap amplitudes are consistent with the expectation $\Delta_{max}/k_B \approx 2T^*$.

## 4. CONCLUDING REMARKS

In conclusion, our analyses of bulk-sensitive crystal-field relaxation experiments on high-temperature superconductors give two important results. Firstly, we gave evidence that the theoretically predicted [20] and for underdoped $Bi_2Sr_2CaCu_2O_{8+x}$ observed [11] gapless arcs of the Fermi surface are also present in the Y- and La-based high-temperature superconductors, thereby being presumably a generic feature of all copper-oxide perovskites. Secondly, there is evidence for an admixture of an appreciable s-wave component of the order of 20-25% to the predominant d-wave gap function in both Y- and La-based high-$T_c$ superconductors. In fact, for orthorhombic symmetry both the s- and d-wave pair states belong to the same irreducible representation [21], thus an admixture of these two states is allowed. Recent angle-resolved electron tunneling experiments in $YBa_2Cu_3O_7$ report similar results, namely a 17% s-wave admixture to the d-wave gap function [22]. Moreover, the gap values listed for the La-based compound in Table 1 are in good agreement with the results of recent in-plane magnetic-field penetration-depth measurements in $La_{1.83}Sr_{0.17}CuO_4$ by muon-spin rotation experiments which gave $\Delta_s$=1.57(8) meV and $\Delta_d$=8.2(1) meV [23]. A subsequent theoretical study confirmed these findings by calculations based on a two-band model [24].

**ACKNOWLEDGMENTS**

The author is very much indebted to K.A. Müller for stimulating discussions.

Table I: Gap parameters resulting from the least-squares fitting procedure applied to the crystal-field relaxation data displayed in Figs. 2(b-d).

| | $T_c$ [K] | $T^*$ [K] | $\Delta_s$ [meV] | $\Delta_d$ [meV] | $\Delta_{max}$ [meV] |
|---|---|---|---|---|---|
| $HoBa_2Cu_3O_{6.56}$ | 60 | 250 | 8.6 ± 1.2 | 28.0 ± 2.1 | 36.6 ± 3.3 |
| $HoBa_2Cu_4{}^{18}O_8$ | 79 | 231 | 9.3 ± 0.9 | 28.5 ± 1.6 | 37.8 ±2.5 |
| $La_{1.81}Sr_{0.15}Ho_{0.04}Cu^{18}O_4$ | 32 | 70 | 2.4 ± 0.4 | 7.7 ± 0.7 | 10.1 ± 1.1 |

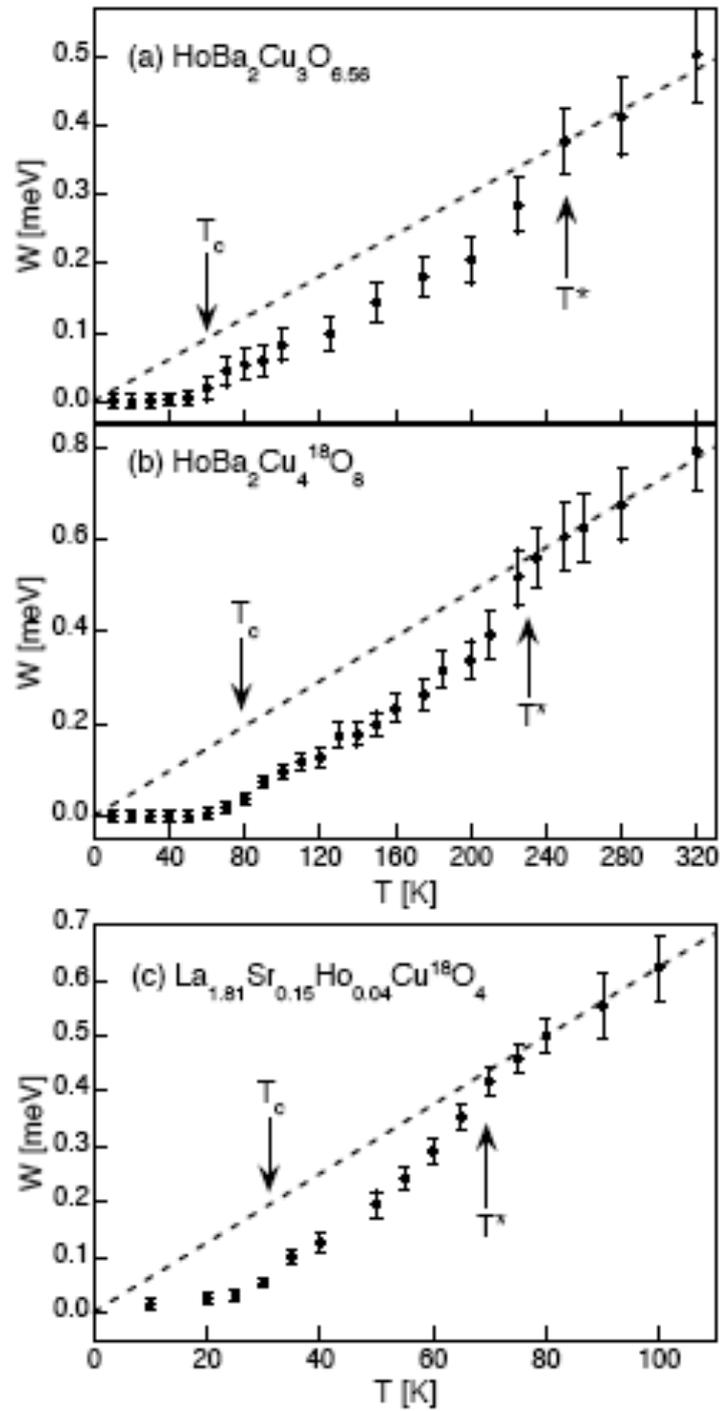

Fig. 1: Temperature dependence of the intrinsic linewidth of the lowest crystal-field ground-state transition in $HoBa_2Cu_3O_{6.56}$ (a), $HoBa_2Cu_4^{18}O_8$ (b), and $La_{1.81}Sr_{0.15}Ho_{0.04}Cu^{18}O_4$ (c). The dashed line corresponds to the linewidth in the normal state as expected from the Korringa law.

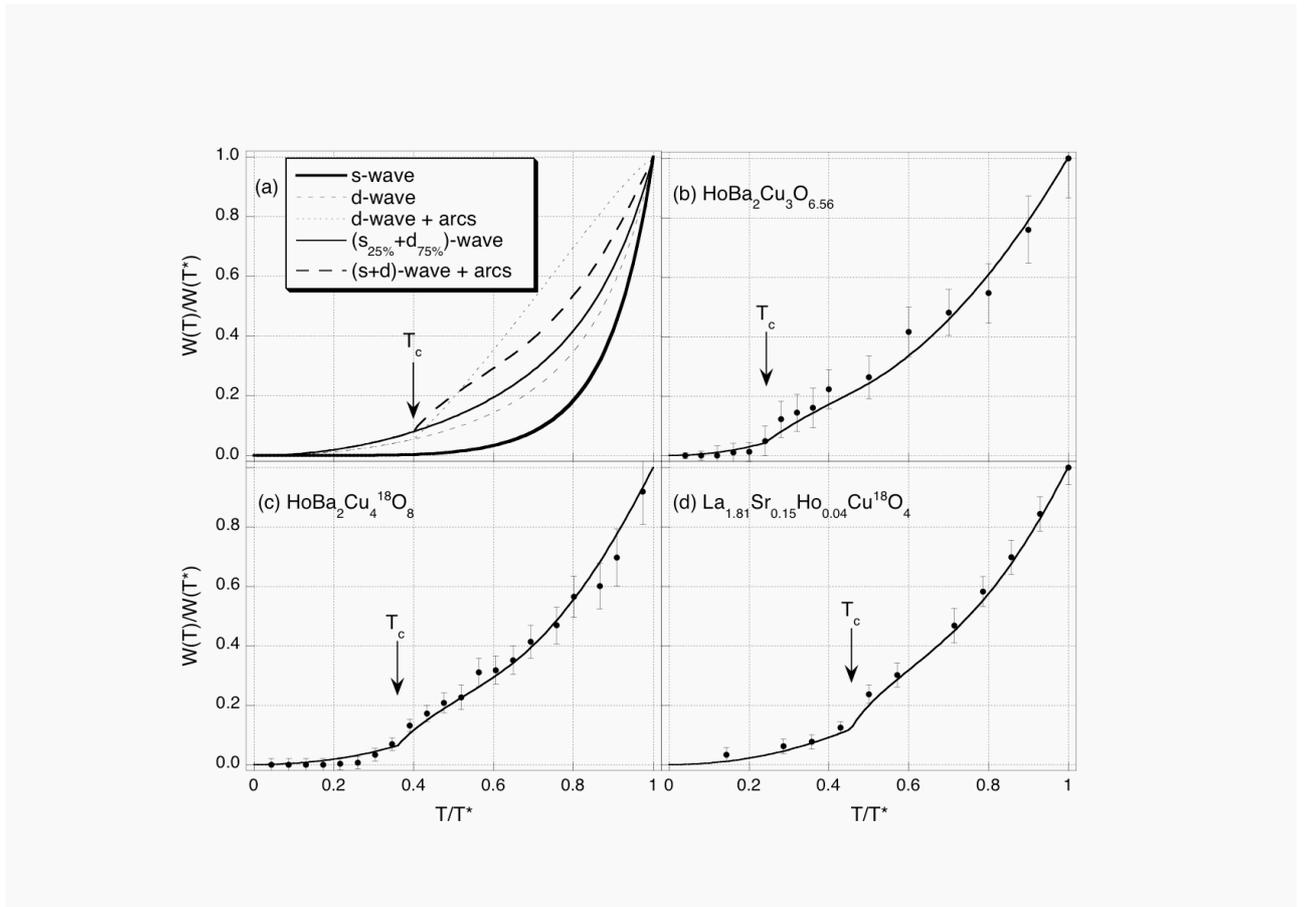

Fig. 2: Temperature dependence of the linewidth of low-energy crystal-field transitions in high-temperature superconductors in reduced units. (a) Model calculations for different types of gap functions (see text). (b-d) Relaxation data derived for $HoBa_2Cu_3O_{6.56}$ (b), $HoBa_2Cu_4{}^{18}O_8$ (c), and $La_{1.81}Sr_{0.15}Ho_{0.04}Cu^{18}O_4$ (d). The lines in (b-d) denote least-squares fits on the basis of a mixed (s+d)-wave gap function including the occurrence of gapless arcs of the Fermi surface in the pseudogap region ($T_c<T<T^*$) as described in the text.

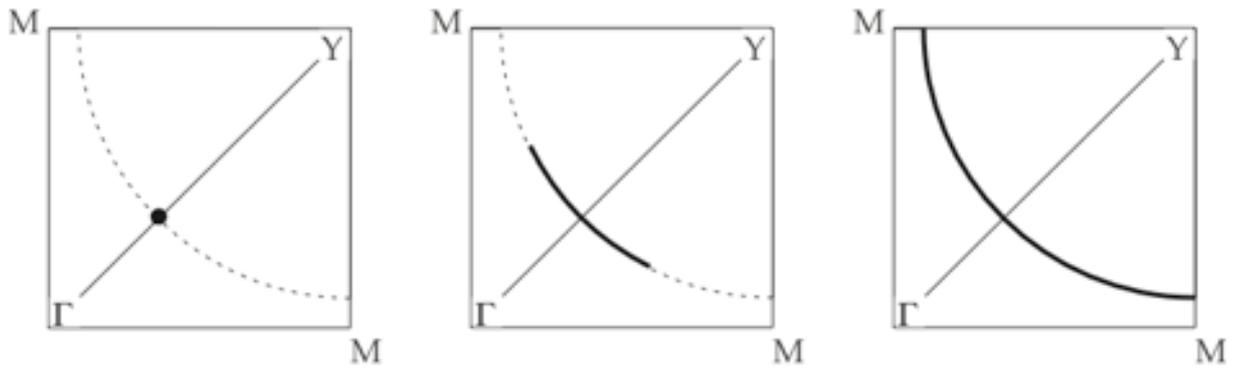

Fig. 3: Schematic illustration of the temperature evolution of the Fermi surface in the pseudogap region ($T_c \leq T \leq T^*$). The d-wave node below $T_c$ (left panel) becomes a gapless arc above $T_c$ (middle panel) which expands upon increasing the temperature to form the full Fermi surface at $T^*$ (right panel).